\documentclass[aps,twocolumn,showpacs]{revtex4}
\usepackage{graphicx}
\usepackage{textcomp}
\usepackage{psfrag}
\begin{document}

\title{Multi-scale Modeling Approach  to Acoustic Emission during Plastic Deformation} 

\author{Jagadish Kumar and  G. Ananthakrishna}

\affiliation{Materials Research Centre, Indian Institute of Science, Bangalore 560012, India}

\begin{abstract}
We address the long standing problem of the origin of acoustic emission commonly observed during plastic deformation. We propose  a frame-work to deal with the widely separated time scales of collective dislocation dynamics and elastic degrees of freedom to explain the nature of acoustic emission observed during the Portevin-Le Chatelier effect. The Ananthakrishna model is used as it explains most generic features of the phenomenon. 
Our results show that while acoustic emission bursts correlated with stress drops are well separated for the type C serrations, these bursts merge to form nearly continuous acoustic signals with overriding bursts for the propagating type A bands.
\end{abstract}

\pacs{83.50.-v, *43.40.Le, 62.20.fq, 05.45.-a}
\maketitle

Acoustic emission (AE)  is observed in an unusually large number of situations. For example, it is observed during crack nucleation and propagation in fracture of solids \cite{Sam92},
micro-fracturing process \cite{Petri94}, martensite transformation \cite{Vives94,Rajeevprl}, peeling of an adhesive tape \cite{Cicc04,Rumiprl}, collective dislocation motion etc \cite{Miguel,Weiss}. Clearly, sources that lead to AE signals in such widely different situations are system specific even as the general mechanism attributed to AE is the abrupt release of the stored strain energy.  The phenomenon is used as a non-destructive tool in understanding the sources and mechanisms generating AE.

Acoustic emission  during plastic deformation refers to high frequency transient elastic  waves generated by abrupt motion of dislocations. AE studies in plastically  deforming  metals and alloys have been reported for over four decades \cite{FL67}.  The changes in the AE signals during deformation  differs from  one type of experiment to another and also  on the sample type. Some correlations has been established between AE signals and the nature of stress-strain ($\sigma-\epsilon$) curves \cite{FL67,CR87,ZR90}.  Conventional yield phenomenon is accompanied by a peak in AE pattern just beyond the elastic regime that decays for larger strains. In contrast distinct AE patterns are observed in the case of unstable plastic deformations such as the L\"{u}ders band and the different types bands in the Portevin - Le Chatelier (PLC) effect  \cite{CR87,ZR90,Ch02,Ch07}. Such differences in AE patterns in different experimental conditions (and samples) can be attributed to the way dislocations respond to external forces. Theoretical approaches to AE are based on Green function approach that use specific  model sources  such as an expanding loop which generate AE \cite{Malen74}.  Clearly, such approaches cannot be useful if one is interested in following the changes in AE  occurring during the course of deformation since AE signals (as also stress) are  averages over dislocation activity in the entire sample. Despite the vast literature on the subject, we are not aware of any  model that predicts the nature of acoustic emission during the entire course of deformation.

The purpose of the paper is to propose a theoretical frame-work to describe both dislocation dynamics and elastic degrees of freedom  simultaneously since it is the abrupt motion of dislocations that transmits the kinetic energy to the surrounding elastic medium triggering the AE signals.  We address the problem in the context of the PLC effect where the signature of the AE signals are  well correlated with the types of serrations and band types observed at different strain rates \cite{ZR90,CR87,Ch02,Ch07}. Our results show that  for type C serrations, the AE bursts which are correlated with stress drops, are well separated. As strain rate is increased, the AE bursts tend to merge to form nearly continuous acoustic signals with overriding bursts for the propagating type A bands. 

It is well known that AE signals in the case of the L\"{u}ders and the PLC bands arise from collective behavior of dislocations\cite{CR87,ZR90,Ch02,Ch07,GA07}. In such cases, it is  necessary to simultaneously describe the collective behavior of dislocations and the elastic degrees of freedom. This so far has not been possible due to several difficulties. First, a major source of difficulty common to all plastic deformation experiments,  is the absence  of theoretical frame work to  simultaneously treat the widely separated inertial time scale and that of dislocation dynamics.  Second, there is lack of dislocation based models to describe collective behavior of dislocations\cite{GA07}.  Third, there is no clarity on how to describe transient acoustic waves. Finally, even in models describing  collective dislocation motion, stress equilibration is assumed. This, however, no longer holds during the process of AE generation \cite{GA07,Bhar03,Anan04}.

The PLC instability is characterized by three types of bands and the associated serrations \cite{GA07}.  On increasing strain rate or decreasing  temperature, randomly nucleated static type C bands are seen, identified with large  stress drops. Then the type B 'hopping' bands are seen. Here, a new band is formed ahead of the previous one in a spatially correlated way giving the visual impression of hopping propagation. The serrations are more irregular with smaller amplitude compared to the type C serrations.   Finally, the continuously propagating type A bands associated with small stress drops are seen.

{\it Our basic idea is to obtain the local plastic strain rate from model equations that describe the entire spatio-temporal evolution of plastic deformation and use it as a source term in the wave equation for the elastic strain.} Here we use the Ananthakrishna (AK) model for the PLC effect \cite{Anan82,Bhar03,Anan04} as it reproduces the band types \cite{Bhar03,Anan04,GA07}, and several  other generic features such as the existence of the instability within a window of strain rates, the negative strain rate behavior etc \cite{Anan82,Rajesh00}. The model also predicts chaotic stress drops which has been subsequently verified \cite{Anan83,Noro97}. The basic idea of the model is that all the qualitative features of the PLC effect emerge from nonlinear interaction of  a few dislocation populations, assumed to represent the collective degrees of freedom of the system. The model consists of densities of mobile, immobile, and decorated (Cottrell) type dislocations denoted by $\rho_m(x,\tau )$, $\rho_{im}(x,\tau)$ and
$\rho_c(x,\tau)$ respectively, in the scaled form.  The scaled evolution
equations are \cite{Anan04}:
\begin{eqnarray}
\nonumber
\frac{\partial \rho_m}{\partial \tau} &=& -b_0 \rho_m^2 -\rho_m\rho_{im} + \rho_{im} -a\rho_m + \phi_{eff}^m \rho_m \\
\label{x-eqn}
&+& \frac{D}{\rho_{im}} \frac{\partial^2 (\phi_{eff}^m(x) \rho_m)}{\partial x^2}, \\
\label{y-eqn}
\frac{\partial \rho_{im}}{\partial \tau} &=& b_0(b_0\rho_m^2 -\rho_m\rho_{im} -\rho_{im} + a\rho_c),\\
\label{z-eqn}
\frac{\partial \rho_c}{\partial \tau} &= &c(\rho_m - \rho_c ),\\
\label{phi-eqn}
\frac{d\phi(\tau)}{d\tau}& = &d[\dot{\varepsilon}_a -\frac{1}{l}\int_0^l \rho_m(x,\tau)\phi_{eff}^m(x,\tau) dx],
\end{eqnarray}
where $\tau$ is the scaled time variable. The term $b_0\rho_m^2$ in Eq. (\ref{x-eqn}),  refers to the formation of dipoles and other  dislocation locks,  $\rho_m\rho_{im}$  refers to the annihilation of a mobile dislocation with an immobile one and the source term $\rho_{im}$ represents the athermal or thermal reactivation of the immobile dislocation. $a\rho_m$ represents the immobilization of mobile dislocations due to aggregation of solute atoms. Once a mobile dislocation starts acquiring solute atoms we regard it as Cottrell-type of dislocation $\rho_c$. As  more and more solute atoms aggregate, they eventually stop, and are considered as immobile dislocations $\rho_{im}$. This is the source term $a\rho_c$ in Eq. (\ref{y-eqn}). $\phi_{eff}^m \rho_m $ in Eq. (\ref{x-eqn}) represents the rate of multiplication of dislocations due to cross slip. This depends on the velocity of  mobile dislocations taken to be  $V_m(\phi) = \phi_{eff}^m$, where $\phi_{eff} = (\phi - h\rho_{im}^{1/2})$ is the scaled effective stress, $m$ the velocity exponent, and $h$ a work hardening parameter.  Further, cross-slip allows dislocations to spread into neighboring spatial locations and thus gives rise to diffusive coupling (last term in Eq. (\ref{x-eqn})). These equations are coupled to  Eq. (\ref{phi-eqn}) that represents the constant strain rate deformation  experiment. In Eq. (\ref{phi-eqn}), ${\dot\varepsilon}_a$ is the scaled applied strain rate, ${\dot\varepsilon}(p,x,\tau) = \rho_m(x,\tau) \phi_{eff}^m(x,\tau)$ is the local plastic strain rate, $d$ the scaled effective modulus of the machine and the sample, and $l$ the dimensionless length of the sample. Note that Eq.(\ref{phi-eqn}) assumes stress  equilibration. The scaled constants, $a,c$ and $b_0$ refer, respectively, to the concentration of solute atoms slowing down the mobile dislocation, the diffusion rate of solute atoms to mobile dislocations and the thermal and athermal reactivation of immobile dislocations. The relevant parameter is the applied strain rate $\dot{\varepsilon}_a$ with respect to which different types of serrations and the associated bands are observed. The instability range is found in the interval $30 < \dot{\varepsilon}_a < 1000$.

Equations (\ref{x-eqn} -\ref{phi-eqn}) are discretized on a grid of $N$ points and solved using a adaptive step size differential equation solver (``MATLAB" `ode15s'). In experiments, bands cannot propagate into the sample due to large strains at the grips. This is mimicked by choosing the boundary conditions  $\rho_{im}(1,\tau)$ and $\rho_{im}(N,\tau)$ to be two orders higher than the rest of the sample. In addition, we impose  $\rho_m(1,\tau) =\rho_c(N,\tau)=0$. The initial values of the dislocation densities are chosen to be  uniformly distributed with a Gaussian spread along the sample. For the numerical work, we use $a=0.8, b=5 \times 10^{-4}, c=0.08, d=6 \times 10^{-5}, m=3.0, h=0, D=0.25, N=100$.  

The above equations (Eqs. (\ref{x-eqn}- \ref{phi-eqn})) are adequate to obtain the plastic strain rate only. However, noting that the abrupt collective dislocation motion triggers the transient elastic  waves, we need to describe both elastic degrees of freedom and dislocation dynamics.  This also implies that instantaneous stress following such an event will display fluctuations that damp-off in course of time. Indeed,  the abrupt slip process induces dissipative forces that tend to oppose the accelerated motion of the slip interface. This is a mechanism that ensures eventual approach to mechanical equilibrium. Following Ref. \cite{Land}, we represent this dissipation in terms of the Rayleigh dissipation function (RDF) given by ${\cal R}_{AE} = {\Gamma\over2}\int\Big[{\partial \dot\epsilon_e(y) \over \partial y} \Big]^2 dy$.  We identify ${\cal R}_{AE}$ with acoustic energy dissipated  by noting that this has the form  of the energy associated with abrupt dislocation motion during  plastic deformation,  i.e.,  ${\cal R}_{AE} \propto \dot \epsilon^2(r)$ \cite{Rumiepl}. Thus ${\cal R}_{AE}$ is taken to be the energy  of the transient elastic waves.  We have shown that the choice of representing the acoustic energy dissipated in terms of Rayleigh dissipation function has been successful in predicting the nature of AE signals in varied situations such as the martensite transformation\cite{Rajeevprl,Kalaprl}, fracture \cite{Rumiepl} and peeling of an adhesive tape\cite{Rumiprl,Jag08a,Jag08b}. Writing down the kinetic energy ($ \frac{\rho}{2}\int (\dot\epsilon^2(y,t) dy$, where $\rho$ is the density), the potential energy ($\frac{\mu}{2}\int \epsilon^2(y,t) dy$, where $\mu$ is the elastic constant), dispersion of the elastic waves  ($\frac{D}{2} [\frac{\partial^2\epsilon_e}{\partial y^2}]^2$ with $D$ a constant), and dissipation ${\cal R}_{AE}$, we get (using Lagrange's equations motion), 
\begin{eqnarray}
\rho \frac{\partial^2 \epsilon_e}{\partial t^2} &=& \mu \frac{\partial^2 \epsilon_e}{\partial y^2} - D \frac{\partial^4 \epsilon_e}{\partial y^4} +\Gamma \frac{\partial^2 {\dot \epsilon_e}}{\partial y^2} - \rho  \frac{\partial^2 \epsilon_p}{\partial t^2} .
\label{comp_wave}
\end{eqnarray}
The second and third terms (on the right hand side) arise from the dispersion and  dissipation terms respectively.  In addition, we have included the plastic strain rate (last term)  calculated from Eqs. (\ref{x-eqn}, \ref{y-eqn}, \ref{z-eqn}), and Eq. (\ref{phi-eqn}). This acts as a source term in the wave equation for the elastic degrees of freedom that is expected to generate  transient elastic waves. Note that Eq. (\ref{comp_wave}) is general and applicable to any plastic deformation situation as long as the plastic strain rate is supplied.   Transforming this equation into scaled variables used in the AK model,  we have 
\begin{eqnarray}
\frac{\partial^2 \varepsilon_e}{\partial \tau^2}= \frac{c^2}{(\theta V_0)^2} \frac{\partial^2 }{\partial y^2}\Big[ \varepsilon_e +\frac{\Gamma}{c^2} \dot{\varepsilon_e}
- \frac{D}{c^2} \frac{\partial^2 \varepsilon_e}{\partial y^4}\Big]-\frac{\partial\dot{\varepsilon}_p(y,\tau)}{\partial \tau}.
\label{waveqn} 
\end{eqnarray}
(The relations between the scaled and unscaled  are $\dot{\epsilon}_k(t)= \frac{bV_0\gamma }{\beta} {\dot \varepsilon}_k(\tau)$  and $\tau = \theta V_0 t$ where $b$ is the Burgers vector, $\beta,\gamma, \theta$ and $V_0$ are constants used in the unscaled AK model equations. See Ref. \cite{Rajesh00} for details.)

Finally, appropriate boundary conditions needs to imposed on Eq. (\ref{waveqn}) that should be consistent with those on Eqs. (\ref{x-eqn}-\ref{phi-eqn}).  This however is not straightforward.  To do this, we first note that numerical solution requires discretization of Eqs. (\ref{x-eqn}-\ref{phi-eqn}) and  Eq. (\ref{waveqn}). Further, as one end of the sample is fixed and a traction is applied to the other end, the total imposed strain rate is shared by the machine and the sample.  This implies that the machine elastic element should be included at both ends, i.e.,  the discrete form of the wave equation should contain equations of motion for the end points of the sample and machine. Then,   the stiffness of the machine enters naturally in the equations for the end points. Then, boundary conditions of Eq. (\ref{x-eqn}-\ref{phi-eqn}) are automatically satisfied by these equations. The  relevant boundary conditions  for  discretized form of Eqs. (\ref{waveqn}) are $\varepsilon_1(\tau) = 0,\,{\rm and} \, \varepsilon_N(\tau) = \dot{\varepsilon}_a \tau$ for $\tau>0$ where the subscript 1 and N refer to the end sites.
The initial conditions are: $\varepsilon_i(0)=0 + \xi,\,\, i=2,..,N-1$ with the random number $\xi$ is drawn from interval $-\frac{1}{2} < \xi < \frac{1}{2}$.

However, the time scale of plastic strain rate (i.e., Eqs.(\ref{x-eqn}-\ref{phi-eqn})) is typically $ \sim {\dot \varepsilon}_a$  while that of Eq. (\ref{waveqn}) is much smaller. Indeed,  the step size in an adaptive step size algorithm  used for the solution of Eqs. (\ref{x-eqn} - \ref{phi-eqn}) are significantly larger that the time step required for integrating  Eqs. (\ref{waveqn}). Thus, we need to ensure that the time variable in Eq. (\ref{waveqn}) and Eqs. (\ref{x-eqn}-\ref{phi-eqn}) are mapped correctly.   Denoting the $i^{th}$ integration time step in the AK model by $\Delta \tau_i$, for the time interval between $\tau_{i+1} < \tau < \tau_i$, we need to ensure that  $ m \delta \tau' = \Delta \tau_i$ where $\delta \tau'$ is the fixed step size used for  Eq. (\ref{waveqn}). Further, we use interpolated values for  the plastic strain rate  $\dot \varepsilon_p(k,\tau)$ (for any $k^{th}$ spatial element)  obtained by using linear interpolation formula  $\varepsilon_p(k,\tau)=\varepsilon_p(k,\tau_i) + \frac{\varepsilon_p(k,\tau_i)-\varepsilon_p(k,\tau_{i+1})}{\tau_i -\tau_{i+1}}\tau$, where $\tau_i < \tau < \tau_{i+1}$ where $\tau_i$ is $i^{th}$ time step of integration of  Eqs. (\ref{x-eqn}-\ref{phi-eqn}). Moreover, the plastic strain rate calculated from  Eqs. (\ref{x-eqn}-\ref{phi-eqn}) has a much  coarser length scale compared to the fine length scale required for wave propagation. Noting that the spatial coupling in the AK model appears only in Eq. (\ref{x-eqn}), it is easy to show that the strain rate  $\dot{\varepsilon}_p(x,\tau)$ in the AK model must be scaled by a factor $\lambda^2$ (assumed to be constant) when used in the wave equation, i.e., $\dot{\varepsilon}_p(y,\tau) = \lambda^2 \dot{\varepsilon}_p(x,\tau)$ where $x$ and $y$ refer respectively to spatial coordinates in the AK model and Eq. (\ref{waveqn}). (The range of $\lambda$ is $10^3 -10^6$.) The results presented are for $N=100,\lambda^2= 10^3, k_m=5k_s,\frac{c^2}{(\theta V_0)^2} =1500,\gamma/(\theta V_0)^2 =10$ and $D/(\theta V_0)^2=1$. Note that the velocity of acoustic waves is of right order for $\theta V_0 \sim 100$.
\begin{figure}
\hbox{
\includegraphics[height=2.8cm,width=4.6cm]{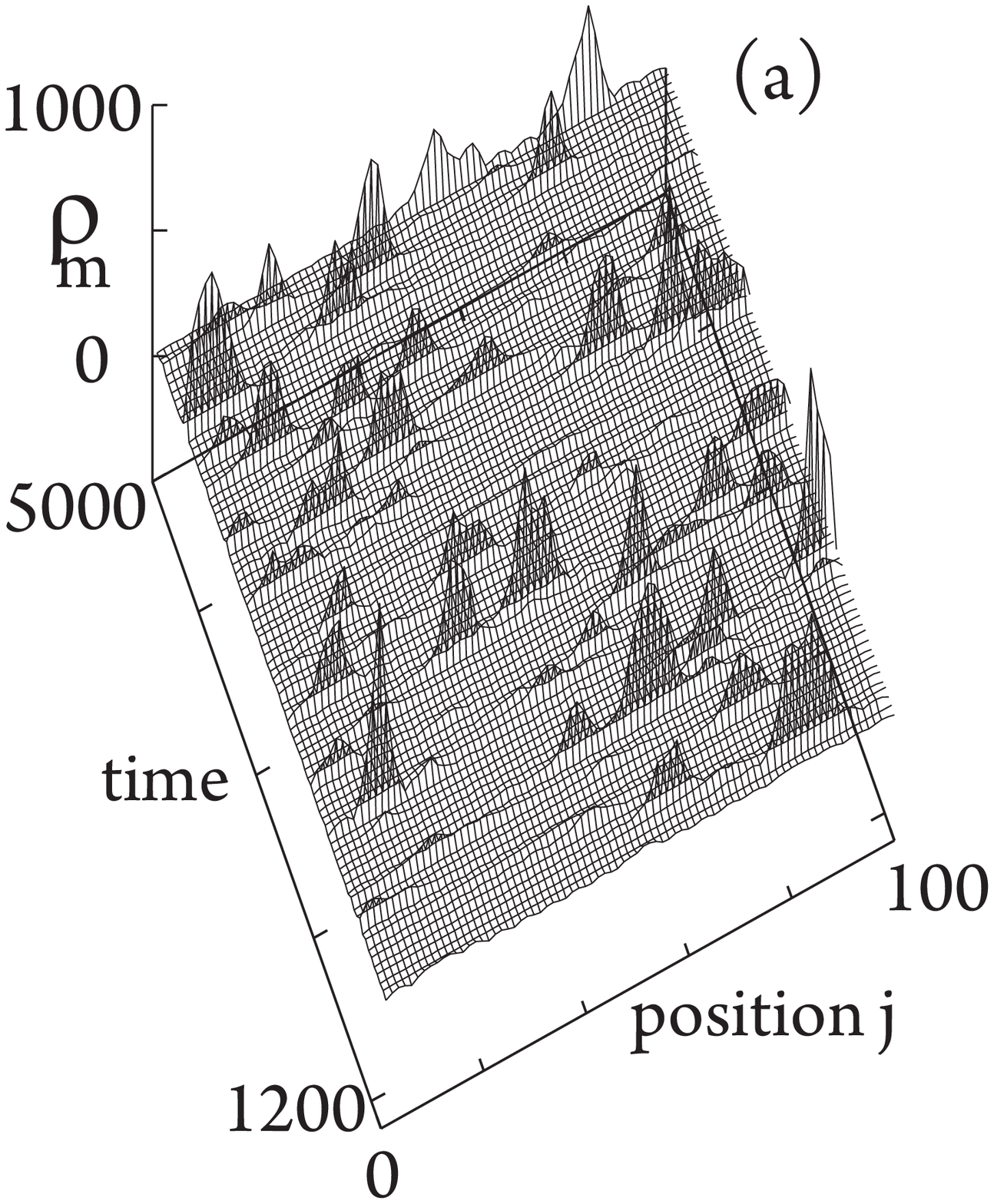}
\includegraphics[height=2.8cm,width=4.2cm]{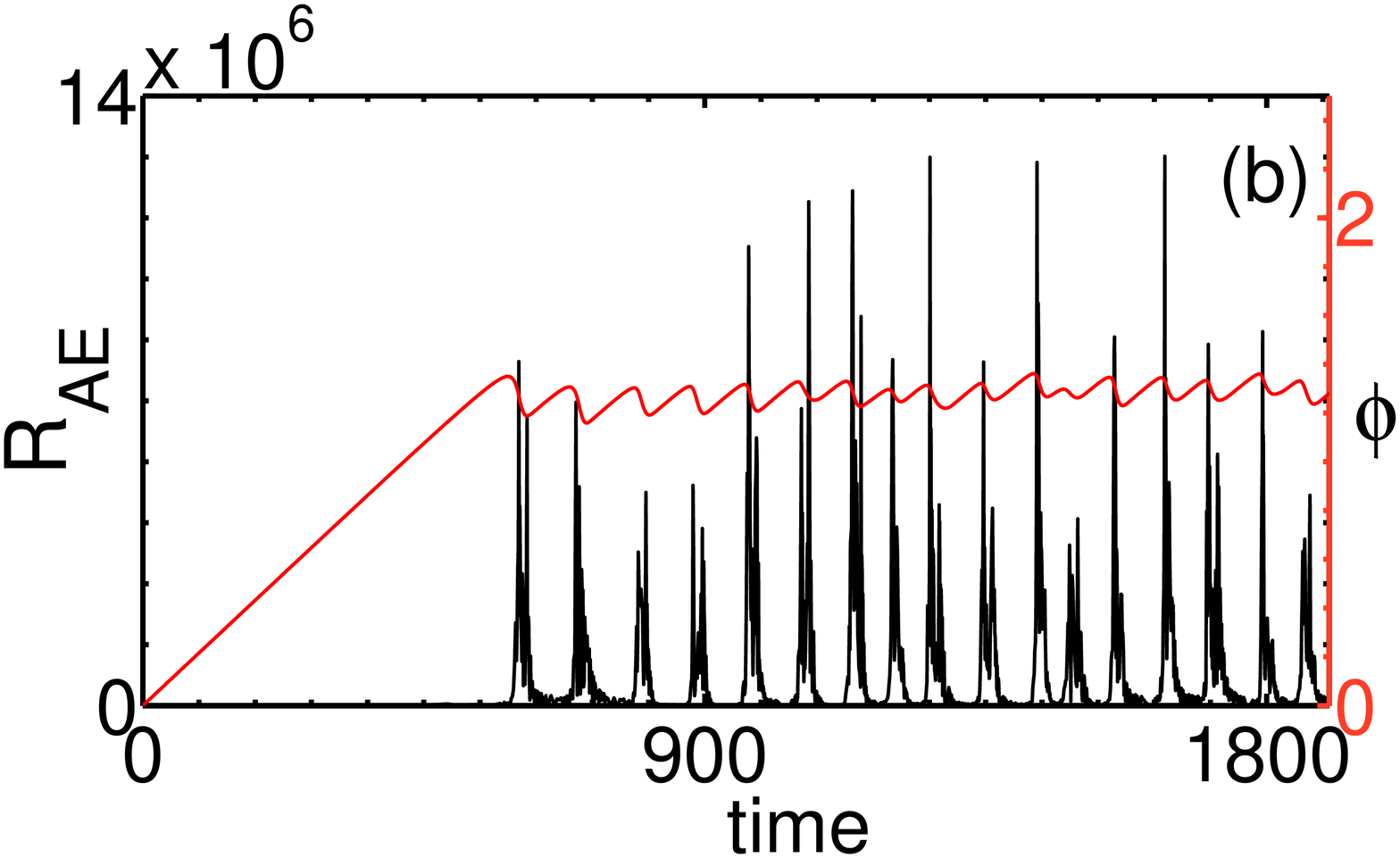}
}
\caption{(a) Uncorrelated type C bands for $\dot{\varepsilon}_a = 40$. (b) (Color online) Plots of stress and acoustic emission energy signals. 
}
\label{PLCAE_BAND_C}
\end{figure}
\begin{figure}
\hbox{
\includegraphics[height=2.8cm,width=4.6cm]{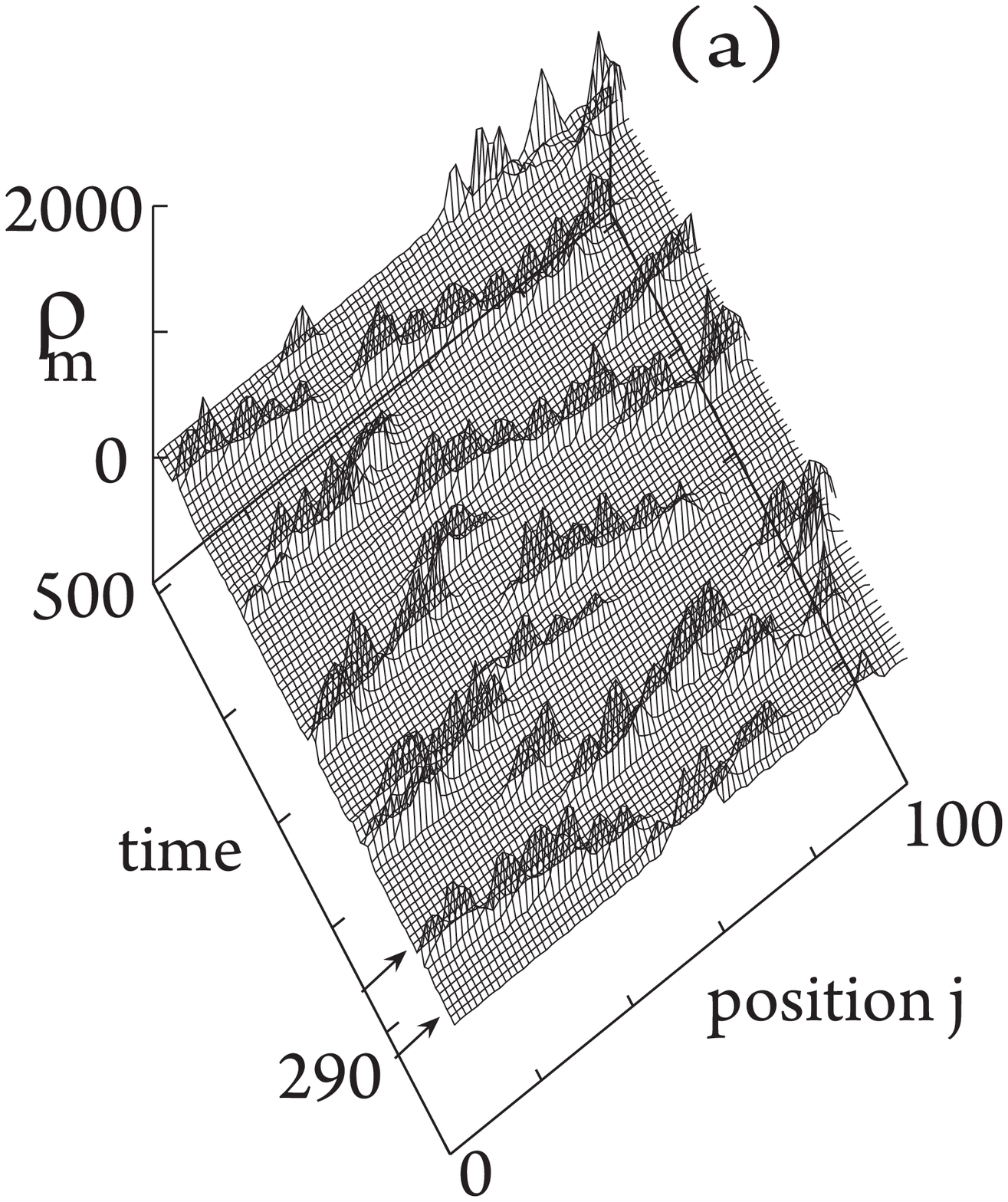}
\includegraphics[height=2.8cm,width=4.2cm]{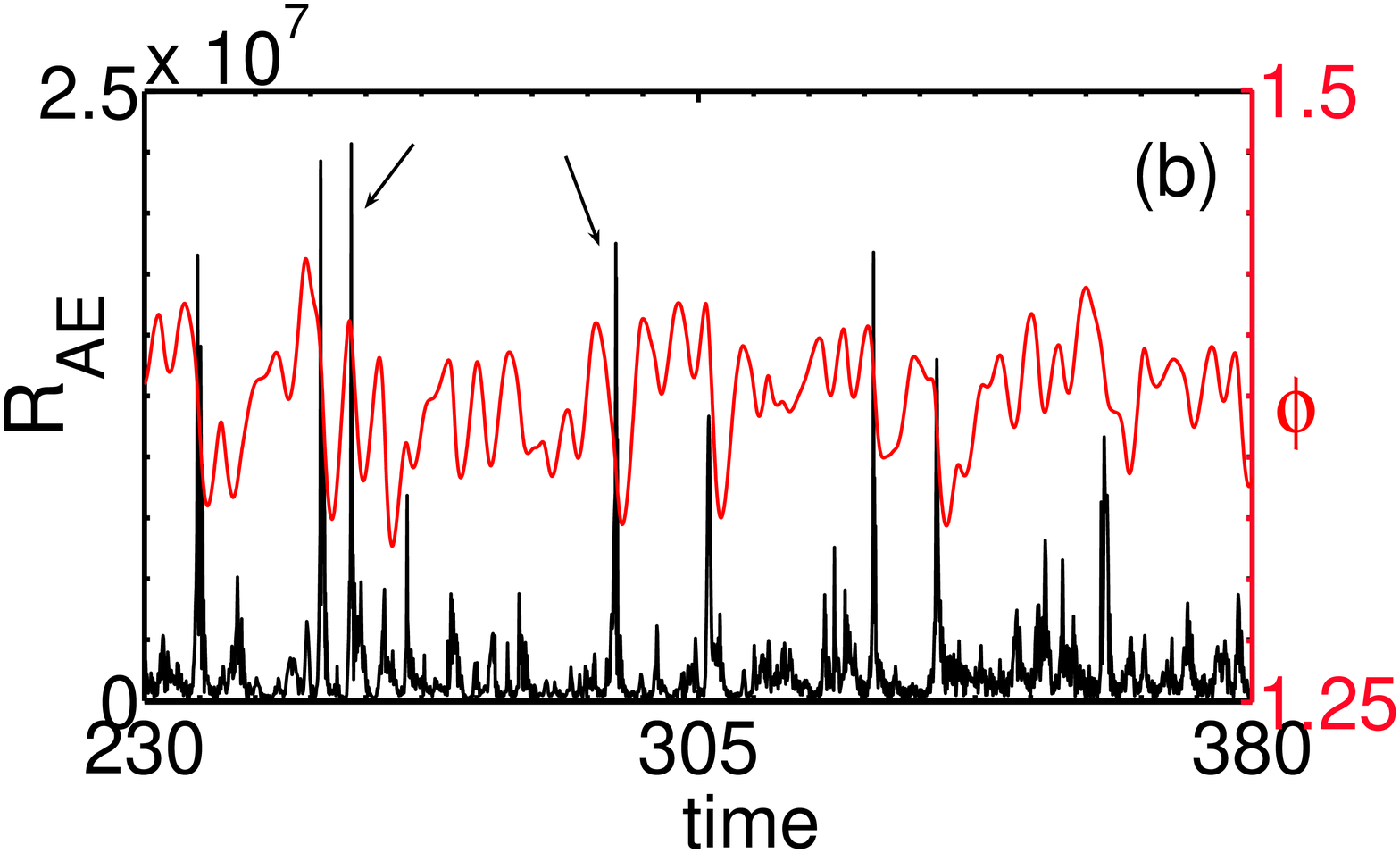}
}
\caption{(a) Partially propagating type B bands for $\dot{\varepsilon}_a = 130$. (b) (Color online) Plots of stress and acoustic emission energy signals in asymptotic regime. The region between the arrows in figures (a) and (b) are identified.
}
\label{PLCAE_BAND_B}
\end{figure}
\begin{figure}[!h]
\hbox{
\includegraphics[height=2.8cm,width=4.6cm]{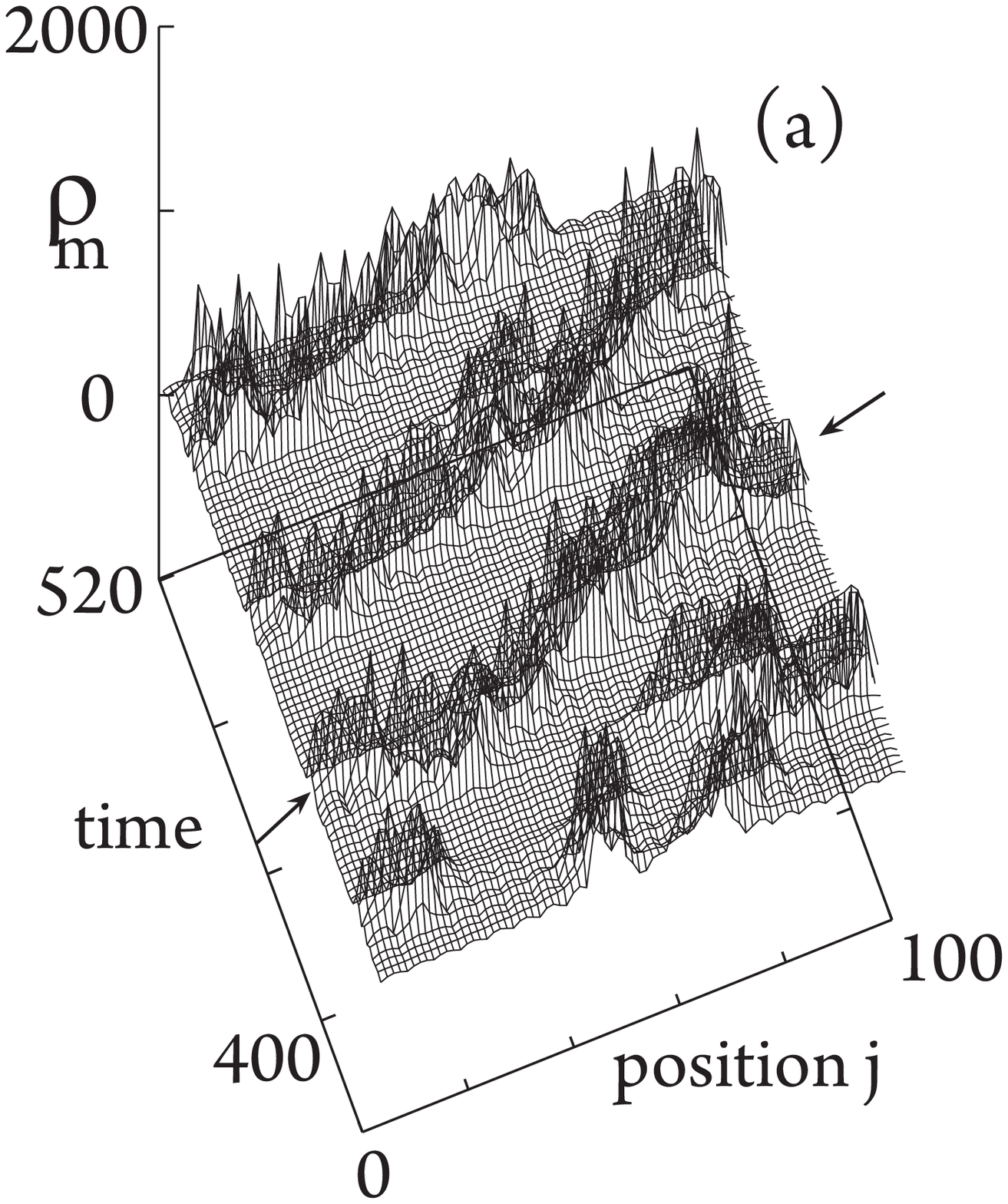}
\includegraphics[height=2.8cm,width=4.2cm]{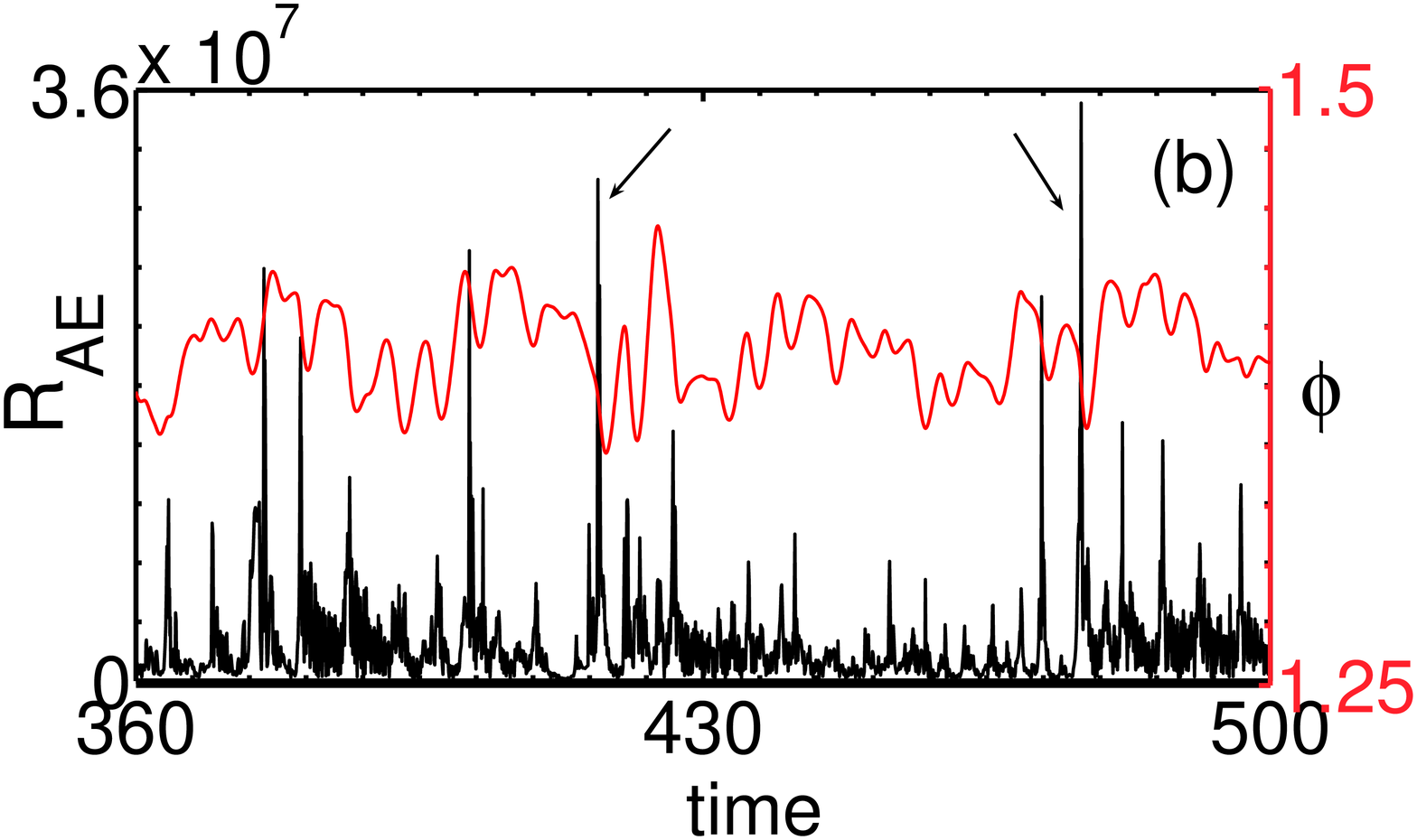}}
\caption{(a) Fully propagating type A bands at  $\dot{\varepsilon}_a = 240$. (b) (Color online) Plots of stress and acoustic emission energy signals in asymptotic regime. The region between the arrows in figures (a) and (b) are identified.
}
\label{PLCAE_BAND_A}
\end{figure}

Equations  (\ref{x-eqn}-\ref{phi-eqn}) and Eq. (\ref{waveqn})  are in principle coupled since ${\dot\varepsilon}_p(y,\tau)$ is a function of stress $\phi(\tau)$.  A self consistent solution of these  equations is equivalent to solving the full dynamical problem involving both plastic deformation and elastic degrees of freedom with the attendant difficulties.  This will not be attempted here. Instead we provide an approximate method akin to adiabatic methods. The procedure adopted  is to first calculate $\dot{\varepsilon}_p(k,\tau)$  for the entire duration of time by solving  Eqs. (\ref{x-eqn}-\ref{phi-eqn}). Then, Eqs. (\ref{waveqn}) is solved   using $\dot{\varepsilon}_p(k,\tau)$ as a source term (along with the scale factor $\lambda$). This gives the elastic strain $\varepsilon_e(y,\tau)$.  Then, the integral of $\varepsilon_e(y,\tau)$ over the specimen dimension gives the transient stress $\phi_{tr}(\tau)$ explicitly. While $\phi_{tr}(\tau)$ will be equal to $\phi$  within the elastic limit, it will be different from $\phi$ beyond this limit. 

The AK model predicts the three band types found with increasing strain rate \cite{Bhar03,Anan04,GA07}.  At low ${\dot \varepsilon}_a$, say ${\dot \varepsilon}_a=40$, the uncorrelated static C bands are seen as shown in Fig. \ref{PLCAE_BAND_C}(a). The serrations are large  and nearly regular.  The scaled acoustic energy dissipated is obtained using $R_{AE} = {\Gamma^s\over2}\int\Big[{\partial \dot\varepsilon_e(y) \over \partial y} \Big]^2 dy$, where $\Gamma^s={\Gamma}/{(\theta V_0)^2}$.  Since, stress drops in this case are due to isolated band nucleation, the AE  pattern consists of well separated bursts that are well correlated with the stress drops.  Figure \ref{PLCAE_BAND_C}(b) shows a typical stress-strain curve along with the AE bursts for $ {\dot \varepsilon}_a =40$.  The post burst AE is continuous that gradually increases until a new burst is seen \cite{ZR90}.

At intermediate strain rates, say  ${\dot \varepsilon}_a =130$ hopping type B bands are seen as shown in Fig. \ref{PLCAE_BAND_B}(a). These propagate partially and  stop  mid-way. Another hopping band reappears in the neighborhood. Often, nucleation occurs at more than one location. The corresponding asymptotic stress-time plot is shown in Fig. \ref{PLCAE_BAND_B}(b). The associated serrations are irregular but are smaller in magnitude compared to the type C.  While the correlation between stress drops and AE peaks still holds when the propagation is short,  the AE bursts are not as well separated as in the case of type C serrations. A plot of the AE signal is shown in Fig. \ref{PLCAE_BAND_B}(b). During hoping propagation, low level AE activity is seen in the region between two AE bursts (shown by arrows) \cite{CR87,Ch02,Ch07}. (A few small bursts are also seen.) As we increase ${\dot \varepsilon}_a$, the extent of propagation increases with concomitant decrease in stress drop magnitudes.  At high $\dot \varepsilon_a$  we find fully propagating type A bands. Figure \ref{PLCAE_BAND_A}(a) shows  dislocation bands nucleating at one end of the sample and propagating continuously to other end for ${\dot \varepsilon}_a =240 $.  The corresponding AE pattern [Fig. \ref{PLCAE_BAND_A} (b)] appears nearly continuous with a few  over-riding bursts. Large bursts in AE are correlated with the nucleation  of the band (or the band reaching the edge or due to occasional intersection of two bands). There is a low level AE activity during propagation (the region between the arrows). In experiments, bands once nucleated trigger a burst in AE but during propagation very low activity is seen \cite{CR87,Ch02,Ch07}. Thus, the generic features of AE signals during the PLC effect are well captured.

In summary, we have developed a theoretical frame-work for dealing with widely separated inertial time scale and that of collective dislocation modes to explain  the nature of acoustic emission patterns observed in the PLC effect. This has been done by computing the plastic strain rate from the AK model for the PLC effect and using it as a source term in the wave equation. An important input in the theory is that the energy of the transient acoustic wave dissipated caused by the  abrupt slip (resulting from collective unpinning of dislocations) is represented in terms of the Rayleigh dissipation function \cite{Rumiprl,Rajeevprl}. The results show that for type C bands, well separated burst type AE signals that are correlated with stress drops are seen. As we increase the strain rate successive bursts tend to merge. For high $\dot{\varepsilon}_a$ where type A propagating bands are seen, the bursts merge to form continuous type of AE signal. Over riding this are AE bursts that correspond to band nucleation or a band reaching the edge. These features are consistent with experimental results \cite{ZR90,Ch02,Ch07}. Other features such as those from hardening can not be captured here as  there is very little hardening in the AK model\cite{Ch02,Ch07}. However, an extension of the AK model that removes this limitation can be used \cite{Ritupan}.  The frame-work is clearly applicable to other deformation conditions as long as a dislocation based model can be developed that captures major features of the  phenomena. 
Finally, better approximate schemes have been designed that also give similar results \cite{Jag10}.

G. A acknowledges the Department of Atomic Energy grant through Raja Ramanna Fellowship Scheme and and INSA for Senior Scientist postion, and also BRNS Grant No. 2007/36/62-BRNS/2564. We thank Prof. A. S. Vasudeva Murthy for useful discussions.

\end{document}